\pdfoutput=1
\documentclass[11pt]{article}

\usepackage{times}
\usepackage{placeins}
\usepackage{graphicx}
\usepackage{latexsym}
\usepackage{amssymb,amsmath}
\usepackage{harvard,ulem}
\usepackage{epstopdf}
\usepackage{enumerate}
\usepackage{bm}
\usepackage{setspace}
\usepackage{fullpage}
\usepackage{sectsty,bbding,lscape}
\usepackage[paper=portrait,pagesize]{typearea}
\usepackage{times}
\usepackage{pdflscape}
\usepackage{anysize}
\numberwithin{equation}{section}
\title{{\sc Measuring Defi Risk}}

\date{}

\author{\\{\sc Jeremy Bertomeu}\hspace{1cm}{\sc Xiumin Martin}\hspace{1cm}{\sc Ibrahima Sall} \footnote{Jeremy Bertomeu is an associate Professor, Xiumin Martin is a Professor and Ibrahima Sall is a Ph.D candidate at the Olin School of Business, Washington University in St Louis, 1 Snow Way Dr, St. Louis, MO 63130.}}

\begin{document}

\begin{doublespace}

\maketitle

\begin{abstract}
Decentralized finance (DeFi) lending has grown from nonexistent in 2017 to nearly 40 billion US Dollars in deposited funds in May 2022. Using cryptocurrency as collateral, the platforms match speculative margin trading with yield-seeking depositors lending coins pegged to the dollar (stable coins). Depositors receive claims guaranteed by a basket of collateral, akin to new stable coins. We develop a framework requiring only knowledge of aggregate deposits and borrowings to measure overall system risks to lenders and borrowers. Using evidence from major protocols, the measures identify an increase in system fragility beyond prudent levels around mid 2021, with a potential loss of peg for extreme variations in coin prices. Overall, the model offers an easily implementable aggregate risk metric capturing the perspectives of synthetic investors and offers early warning signals as the industry is moving from deposits guaranteed by collateral to fiat money.
\vspace{0.3in}

%\vspace{0in}\\
%\noindent\textbf{JEL Codes:} E42, E51, G21, G23, G32, G33.\\
\bigskip
\end{abstract}

\newpage

\paragraph{}

\begin{figure}[ht]
\begin{center}
\includegraphics[scale=1.2]{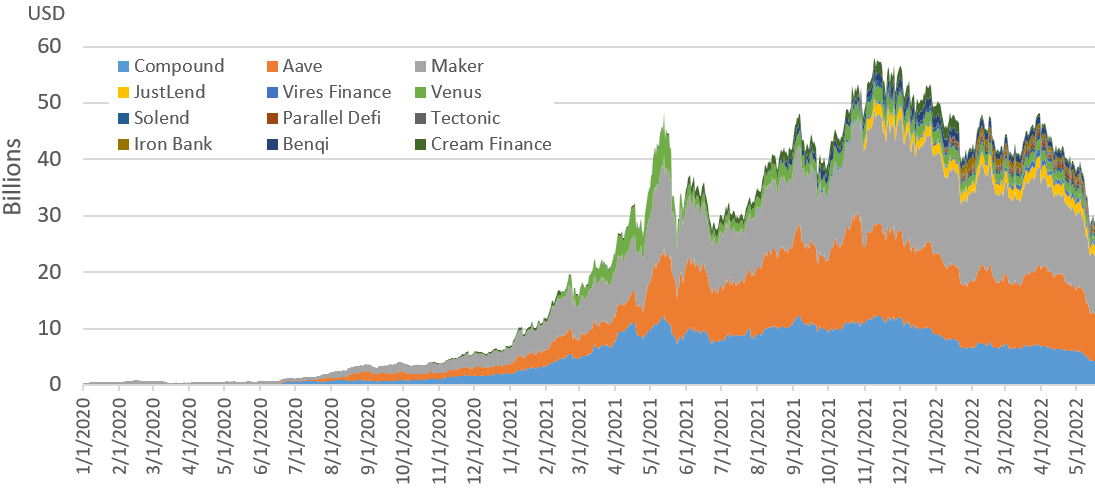}
\caption{Sample of largest Defi lending protocols\label{fig1x}}
\end{center}
\end{figure}

\paragraph{}
The market for cryptocurrency loans has exploded over the recent years. Decentralized finance (DeFi) loans are the highest growing segment in this industry, totalling a total deposited value of nearly 40 billion USD in May 2022, from roughly 500 million USD in January 2020, see Figure \ref{fig1x}. DeFi lending allows owners of cryptocurrency assets to earn interest by depositing their assets into a lending pool, while borrowers access credit from the pool subject to depositing collateral.\footnote{See \citeasnoun{meywelsan21} for a review of the recent literature.} Normally, lenders deposit stable coins (pegged to the dollar) in exchange for interest payments while borrowers borrow the coins, sell them in exchange for unpegged coins such as bitcoin or ethereum, depositing an amount greater or equal in unpegged coins into the pool as collateral.

The resulting lending pool is a margin investment account  using collateralized debt in stable coin guaranteed with a sufficiently large reserve of unpegged coin. Lenders own tokens with a claim to recover their deposits, potentially converting deposits in stable coins guaranteed by audited dollar balances (e.g., USDC) into new stable coins in the form of deposit tokens in the platform guaranteed by a basket of cryptocurrencies (e.g., cUSDC). The recent collapse in the algorithmic coins Terra-UST, with a circulating supply of 12 billion USD, has drawn renewed attention to the stability of tokens without dollar reserves. However, the risks of DeFi deposit tokens, which share some of the stabilization principles of algorithmic stable coins (i.e., they are guaranteed by a basket of cryptocurrencies), are not fully understood.

\FloatBarrier
 
 \FloatBarrier
 The objective of this article is to clarify the market structure matching lenders and borrowers, and quantify the risk of the pool as a function of downward changes in the value of coins. We develop a simple framework to assess the position of a pool relative to lenders and borrowers and derive two system risk measures: (i)\ the risk of liquidation for a synthetic margin borrower and (ii) the risk of collateral loss for a synthethic lender.
The measures are implementable with aggregate public data on the borrowing and lending positions of major platforms without knowledge of investor-level leverage.\footnote{Unfortunately, measuring investor-level leverage is difficult given the anonymity of wallets.\ Although wallet addresses and their content are public on the ledger, some wallets may represent exchanges and large investors have multiple wallets, making it infeasible to consolidate  net positions at the investor level. This data limitation motivates the approach of this paper to measure risk at the institution level in terms of a synthetic lender matched to a synthetic borrower.} 

The growing popularization of cryptocurrencies among investors presents several new challenges in the DeFi segment. DeFi deposits are perceived as products with low risk of default, because borrowers must submit more than the value of the loan in collateral and will be liquidated if the ratio of collateral cannot be maintained. Further, while platforms report deposits in each pool, the net position of a pool at any point in time, which can be a complex function of smart contracts linked to the pool, is not easily accessible. Unlike a traditional money fund, lending pools do not report a net asset value. Many depositors may face exposures to large declines in coin prices that would cause the value of collateralized assets in the pool to fall below the claims of depositors.

To set ideas, we describe below the characteristics of a DeFi loan. This type of loan is different from a traditional bank loan, where a lender deposits funds at the bank, receiving in exchange a claim on the bank (or bank account) and the bank lends the funds to a borrower, receiving in exchange a claim on the borrower. This system is centralized because the lender and the borrower transact directly with the bank. By contrast, the objective of a DeFi loan is to match directly the borrower and the lender, automating the functions of the bank via a self-executing (or smart) contract.

This of course presents some challenges. Unlike a bank, a smart contract cannot evaluate soft information to measure credit risk. Therefore, DeFi loans are structured to dynamically manage risk of default. At initiation, the borrower must deposit more than the value of the loan as collateral, resulting in many cases in a ratio of a minimum of \$1 of deposited assets for \$.8  of loan, where $.8$ is the loan-to-value ratio and can vary across types of collateral. Most platforms normalize this ratio by multiplying by the liquidation threshold, denoting it the health factor: a borrower must maintain a health factor greater than one. Otherwise, any outstanding loan is open to a decentralized liquidation process, where an investor acting as liquidator can  use the collateral to repay the loan in exchange for a liquidation bonus.\ The collateral ratio and bonus vary across collaterals and tends to be higher for more volatile assets. 

To our knowledge, prior research on the risks of DeFi lending is extremely limited and, given the empirical challenges it poses and the novelty of the industry, there is no systematic assessment of the positions taken by lenders and borrowers. Notable exceptions in computer science are \citeasnoun{gudperhar20},  \citeasnoun{perwerxu21} and the commercial Gauntlet simulation platform by \citeasnoun{Kao2020AnAO}. However, the focus of these studies is different from ours. They aim to perform detailed stress tests at particular point in time based on based on an agent-based model of contagion, consistent with \cite{sch20}, across cryptocurrencies and requires substantial inputs from a DeFi protocol. Our objective, by contrast, is to develop a measure that can be easily tracked in a time-series with aggregate data and may point to a need to conduct detailed stress tests if positions become sufficiently risky.  

Our study connects to a growing literature examining the 
risk and returns of Fintech products. Several white papers highlight the risks of DeFi functionalities, not just the financial risk to lenders, see, e.g., \citeasnoun{allengujagtiani21} and \citeasnoun{harramsan21}, calling for more research in this area. Studies in this area are dominantly related to earlier Fintech products and services, such as initial coin offerings  \cite{simzya19,bourveau22}, Fintech mortgage instruments \cite{bucmatpis18,agarwalzhang20}, or P2P lending \cite{Rantanphankeppo19,Hsulibao20}. P2P lending and Defi lending are both substitute to traditional forms of finance and rely on online platforms that connects borrowers and lenders. However, while P2P lending is based on traditional currencies supplemented with borrower screening, DeFi lending is based on over-collaterization with a portfolio of cryptocurrencies.

\section{Conceptual Framework}

\label{DeFi2}
\paragraph{}
We develop  a simplified conceptual framework that will serve to lay out the key elements of DeFi loans. Consider two trading dates $t=0,1$, a lending pool and two traders A and B, a price path $(p_t)_{t=0,1}$ for ETH  with initial price $p_0=1$ normalized to one dollar.
The borrowing (lending) rates on ETH $k_{t}^e$ ($v_{t}^e$) and USDC $k_{t}^u$ ($v_{t}^u$).
USDC is a stable numeraire coin with price of $\$1$ in both periods \cite{catgor21}.
Trading ends at the end of period 1 and the two traders convert their holdings into numeraire.  

At date $0$, trader A initially owns $w_A$ USDC and trader B owns $w_B$ ETH. Trader A is solely interested in yield, while trader B wishes to leverage a position in ETH.\ Trader A deposits $w_A$ USDC into the lending pool in exchange for token tUSDC that can be redeemed for USDC. Note that tUSDC is a stable coins which, unlike USDC (which is guaranteed by an audited dollar deposit), is guaranteed by the basket of securities committed to the protocol. Trader B deposits $w_B$ ETH as collateral and borrows $\theta$ USDC at a rate $k_{t}^u$ and exchanges them for $\theta$ ETH. The purchased ETH is deposited in the pool to obtain additional interest-bearing tokens and can also be used as collateral.
 There is a liquidation threshold $\tau^b$ on ETH so that, to maintain a health factor above one, the ratio of collateral to borrowings must satisfy
\begin{equation}
h_0\equiv \frac{w_B+\theta}{\theta}\tau^b\geq 1. 
\end{equation}

Most DeFi platforms use a variation on the following model to set borrowing and lending rates. The borrowing rate is a preset increasing convex function $k_t^x\equiv \phi_x(U_t^x)$, where we denote $x=e$ for ETH loans and $x=u$ for USDC loans and the utilization rate $U_t^x$ is the ratio of borrowings to deposits in currency $x$. The ``base rate" if no trader is borrowing is zero, so we set $\phi_x(0)=0$. It follows that the borrowing interest on USDC\ loans is $k_0^u=\phi_u(\theta/w_A)$, while the borrowing interest on ETH loans is $\phi_e(0)=0$.

The interest paid to lenders can then be recovered by equating the inflow of interest paid by borrowers to the outflow of interest by lenders.\ In the case of USDC\ loans,
\begin{equation}
w_A v_0^u-\theta k_0^u=0,
\end{equation}
that is, substituting the interest on USDC loans $k_0^u=\phi_u(\theta/w_A)$, the interest paid to lenders on the USDC loan is
\begin{equation}\label{lend1}
v_0^u=\frac{\theta \phi(\theta/w_A)}{w_{A}}.
\end{equation}
Unlike in traditional lending, the spread $k_0^u-v_0^u$ does not represent a commission earned by an intermediary. DeFi protocols do not set rates that equate supply and demand, and there are usually more lenders than borrowers. The spread reduces the rate received lenders if there are fewer borrowers, as if borrowers were to randomly pick a lender.

Consider next the ETH\ price $p_1$ at date $t=1$. The health factor $h_1$ of trader B is now given by
\begin{equation}\label{h1}
h_1\equiv \frac{(w_B+\theta)p_1}{\theta(1+\phi(\theta/w_A))}\tau^b.
\end{equation}
There are two possible scenarios:

\begin{enumerate}
\item[1.] If the health factor in (\ref{h1}) remains greater than one,  trader B's health factor is adequate and no collateral is put for liquidation. The loan is fully repaid with traders A and B achieving ending balances $w_A'=w_A+\theta\phi(\theta/w_A)$  and $w_B'=(w_B+\theta)p_1-\theta(1+\phi(\theta/w_A))$, respectively.

\item[2.] If the health factor falls below 1, liquidators can repay up to 1\ USDC, and receive in exchange $(1+\pi)/p_1$ ETH in  collateral per unit of repayment, where $\pi$ is a ETH liquidation premium fixed in the protocol. This implies two subcases:

\item[2.a.] If the loan and premium can be paid in full with the collateral, that is, 
\begin{equation}\label{EQ1}
(w_B+\theta) p_1\geq \theta(1+\phi(\theta/w_A))(1+\pi),
\end{equation}
trader B is liquidated and keeps only the remaining collateral. Trader A achieves the same terminal  wealth as in scenario 1 while trader B achieves a lower $w_B'=(w_B+\theta)p_1-\theta(1+\phi(\theta/w_A))(1+\pi)$ reduced by the premiums paid to liquidators.\footnote{Many DeFi platforms have scripts that allow the depositor to automatically repay the loan once the loan becomes eligible for liquidation, thus avoiding the premium $\pi$.\ With these ``voluntary" liquidation calls, the loan is fully repaid and scenario 2a is equivalent to scenario 1.}

\item[2.b.] If the price of ETH $p_1$ falls more so that the loan can no longer be fully repaid, i.e., inequality (\ref{EQ1}) no longer holds, liquidators can repay $x$ units of USDC loan (cum interest) and claim up to the available collateral. Assuming that liquidators claim the maximum available collateral, the total loan repayment will be
\begin{equation}
(1+\pi)x=(w_B+\theta) p_1,
\end{equation}
implying that $\theta(1+\phi(\theta/w_A))-\frac{w_{B}+\theta}{1+\pi}p_1$ remains  unpaid to the lending pool. After impairment, trader A converting the tokens can recover 
\begin{equation}
w_A'=w_A-\theta+\frac{w_{B}+\theta}{1+\pi}p_1<w_A+\theta\phi(\theta/w_A),
\end{equation}
while trader B loses their entire investment.

\end{enumerate}

Note that the tokens of trader A in the lending pool are not risk-free if scenario 2b has non-zero probability. The lending pool has a standard debt-like risk and return profile: constant for prices of ETH satisfying (\ref{EQ1}) but, otherwise, if trader\ B defaults, tracking the value of a portfolio of ETH\ and USDC.
The critical price level 
\begin{equation}\label{uP}
\underline{p}_1=\frac{\theta(1+\phi(\theta/w_A))(1+\pi)}{w_B+\theta}
\end{equation}
where (\ref{EQ1}) is met at equality, i.e., such that the net assets in the pool are no longer sufficient to guarantee the claim of trader A, are the basis of our risk measure.

\section{Risk\ Measures}
\label{DeFi3}
\paragraph{}

We are now equipped to state the general model and assess the risk of the system. There are two trading periods $t=0,1$ and $j\in [-J,J]$ coins. We denote $j\leq 0$ as stable coins whose value is pegged to USD. Each  coin has an initial price normalized to one, and value at $t=1$ is denoted $p_j$. There are $i\in [1,I]$ traders with a date $t=0$ portfolio given by $\theta_{ij}$. A portfolio $\theta_{i,j}<0$ indicates that trader $i$ is short on coin $j$. Each coin has its own liquidation threshold $\tau_j$ and premium $\pi_j$. Given that we will be interested in instantaneous risks in the system (one or two days), we omit interest rates as they are negligible on a daily basis relative to price changes. 

The total deposits $D_j$ and borrowings $B_j$ in coin j are given by:
\begin{eqnarray}
D_j=\sum_{i=1}^I 1_{\theta_{ij}\geq 0}\theta_{ij},
\hspace{1cm}B_j=\sum_{i=1}^I 1_{\theta_{ij}<0}|\theta_{ij}| .
\end{eqnarray}

Unfortunately, the true risk of the system, defined as the risk that deposits are unpaid, depends on many factors that are unobservable, such as the joint distribution of future coin prices and individual net position as well as leverage on investor-level margin accounts. To address this, we develop two synthetic measures of risk that can be computed with available aggregate deposits and borrowings and can be interpreted in terms of a general coin price level.

The measures are designed to capture risk from the perspective of the two types of 
synthetic traders described in section \ref{DeFi2}: type A seeks yield by investing only in the $j\leq 0$ stable coins while type B seeks capital appreciation by borrowing the $j\leq 0$ stable coins and trading the remaining $j>0$ coins. Note that these are not meant to be a literal representation of a trader but an aggregate of the two main motives for using\ DeFi.
Type B is an aggregate representation of many smaller traders with different levels of leverage who, individually, might not purchase the market portfolio of all unpegged coins. In practice, some traders may be combinations of the types, seeking a mix of yield and coin appreciation.\ As we consider a risk measure for each type of synthetic trader, a mixed trader would have a risk at an intermediate level between the two risk measures.

Hereafter, we operationalize risk as a proportional reduction in all unpegged coin prices that would cause a specific negative event (to be defined later on). A risk measure of zero means that coin prices would have to fall to zero value for the event to trigger, while a risk measure of one or above means that coin prices at the current level ($100\%$ of current prices) trigger the event. 

The first risk measure captures a liquidation event by the type B trader. Assuming that coin prices decrease by a factor $p_I$, type B's health factor is 
\begin{equation}\label{RI1b}
h\equiv  \frac{p\sum_{j> 0} (D_j - B_j)\tau_j}{\sum_{j\leq 0} B_j},
\end{equation}
where the numerator is defined by the net wealth after the price change weighted by the collateral requirement $\tau_j$. 
Our interest is in the maximum price change such that the health factor falls below one. Solving (\ref{RI1b}) in $p$ after setting $h=1$, the first risk measure is defined by
\begin{equation}\label{RI1}
        R_{I}\equiv\frac{\sum_{j\leq 0} B_j}{\sum_{j> 0} (D_j - B_j)\tau_j}
\end{equation}
and captures the overall price change that can lead to liquidation by the synthetic type B\ investor.

From the perspective of the type A\ depositor, most liquidations need not imply default because the loans are over-collateralized with $\tau_j\in (0,1)$. Hence, we consider a second measure capturing the price change such that type B trader is no longer able to repay the loan. After a forced liquidation, the value of the portfolio of type B trader is    
\begin{equation}\label{TB}
p\sum_{j>0} (D_j-B_j)- \sum_{j\leq 0} B_j(1+\pi_j),
\end{equation}
where the first term is the value of the portfolio prior to liquidation and the second term is the repayment amount plus bonus to liquidators. Solving for $p$ so that (\ref{TB}) remains positive,
\begin{equation}
R_{II}\equiv \frac{\sum_{j\leq 0} B_j(1+\pi_j)}{\sum_{j>0} (D_j-B_j)}
\end{equation}
is the point at which type B may not longer be solvent and, therefore, loans may become under collateralized.

Note that this decomposition is different from the common approach which consists in measuring the overall ratio of borrowing to lending to assess the leverage of a synthetic investor. This approach implies by construction a ratio lower than one and may be problematic because lenders and borrowers are different traders with different risk exposure. Put differently, even if lending is significantly greater than borrowing at a point in time, such lending would drain quickly should there be collateral losses in the lending pool: our risk measures attempt to capture aggregate excess collateral deposited by borrowers against commitments to lenders.

\section{Empirical Analysis}

\paragraph{}
We compute the measures daily from April 1st 2021 to May 18th 2022 for Compound and AAve, for a total of 27,804 daily token-level observations.\footnote{These protocols experienced major changes in 2020 and early 2021 due to the opening of new markets and changes to
details of the protocol in response to stress events (e.g., available markets, coins available for trading, and risk parameters). For
this reason, we start the sample after protocols became more homogenous. For example, early 2021, protocols altered their oracle prices and updated liquidation procedures to address events with low liquidity or risks of price manipulation. For simplicity, we do not incorporate the other large DeFi protocol Maker. Maker functions with a different model: borrowers deposit assets in exchange for a stable coin (DAI) that they can sell on the market, so that lenders are owners of the DAI coin.} The Aave and Compound protocols are among the largest in DeFi lending. Detailed information about lending and borrowing is obtained from the Ether blockchain from the `c' tokens issued by Compound and `a' tokens generated by AAve (e.g., aUSDC or cUSDC). These tokens are issued when funds are deposited into the protocol.\footnote{Queries are publicly available at https://dune.com/queries/[query number], as queries 462944, 462941, 463938 and 462905. These queries are available to use and modify, with reference to the current study.} We do not include as stable coins any coin backed exclusively by a portfolio of other unpegged cryptocurrencies, which is consistent with the definition in the new proposed U.S. legislation S.3970 on stable coins. Further, our purpose here is to evaluate the risk of non-repayment of an asset whose value is exogenous to the crypto currencies, so this classification rules out coins supported by a basket of coins such as Terra USD or DAI.\footnote{On June 17th 2022, Maker suspended the DAI Direct Deposit Protocol (D3M), which allowed the minting of new DAI to offset the increase in interest rate given additional borrows in DAI on the platform. This decision was in response to about 100 million DAI having been borrowed by Celsius, a platform currently facing severe liquidity issues.}

We find that from 2021 to 2022, the risk measures became briefly elevated in June and July 2021, where, at peak, $R_{I}$ was greater than one and an additional loss of roughly $1-R_{II}\approx 35\%$ in coin prices would have compromised the deposited collateral. DeFi deposits consolidated over the last six months, with the risk measure $R_I$ indicating that a crash of at least $1-R_I \approx 30\%$ would compromise the position of borrowers. We conclude that, Defi positions had historically been at high levels of fragility and that at current levels, the risk is moderate relative to daily variation in coin prices. However, Defi positions remain vulnerable to a crash or a return to pre-Covid crypto valuations. Note that the risk measures allow for continuous monitoring of this exposure.

\begin{figure}[ht]
\begin{center}
\includegraphics[width=0.59\linewidth]{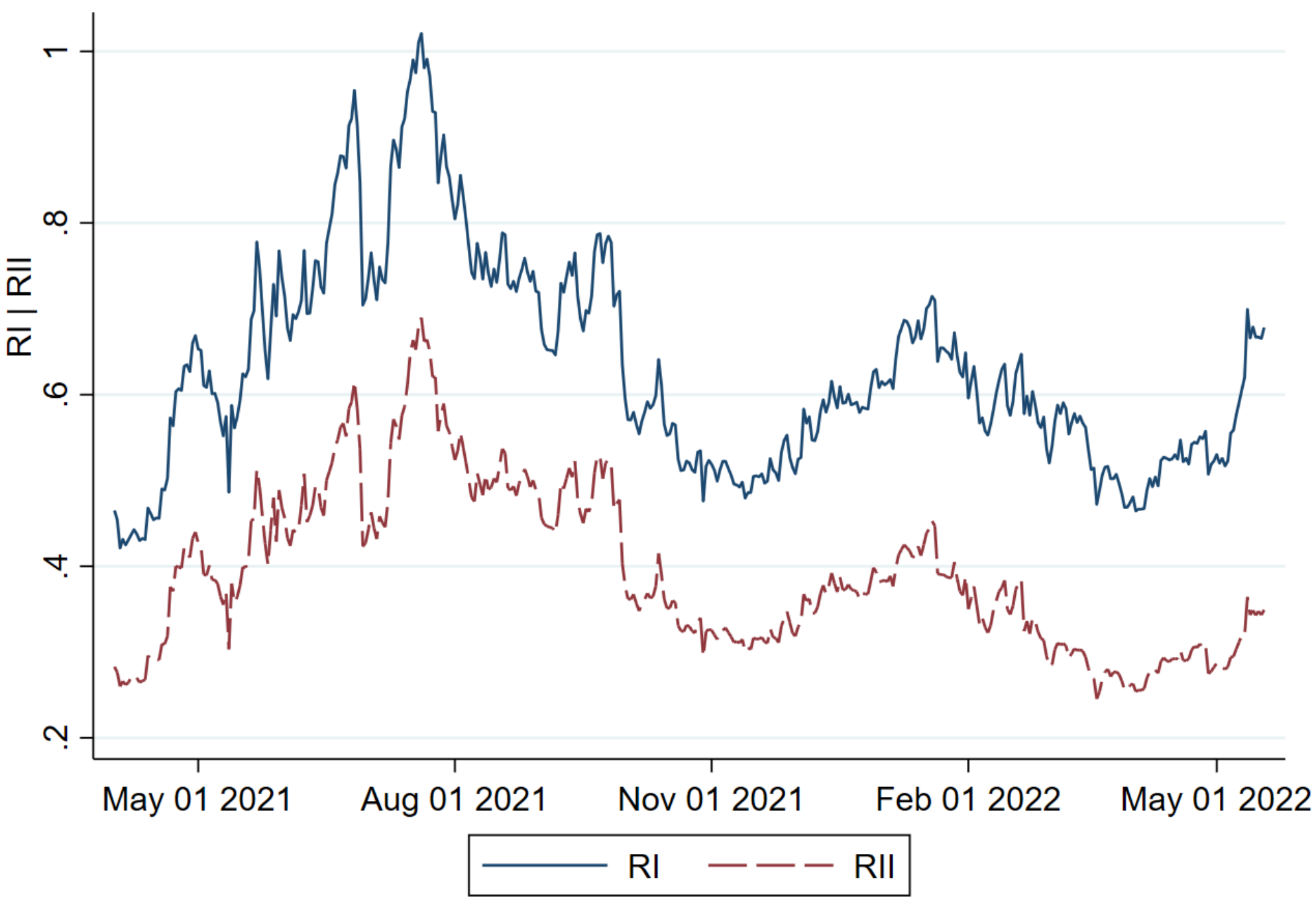}
\caption{Plots of computed $R_{I}$ and $R_{II}$ - Daily}\label{fig5x}
\end{center}
\end{figure}
Next, we explain changes to risk in terms of events relevant to the market for cryptocurrency as well as external financial and macroeconomic shocks. To this effect, we use the general performance of a crypto currency basket, macroeconomic factors, such as inflation and unemployment, and financial factors, such as the yield curve and the value-weighted equity market. The statistical model is stated below:
\begin{equation}\label{r1}
                R_{i}^t =\beta_{0,i}+ \beta_{1,i} \textrm{NCI}_t+\textrm{Controls}+ \epsilon_{i,t},
\end{equation}
where $R_i^t$ denotes $R_{I}$ or $R_{II}$. The explanatory variables are NCI, the Nasdaq value-weighted level of the coin index NCI. The control variables include: inflation, unemployment rate, gold price, spread between the 3-month and the 10-year treasury yields, level of the CRSP value-weighted equity index, Chicago Board Options Exchange volatility index, and Schwab inflation-indexed
bond. Our estimates, presented in table \ref{table5x}, suggest that an increase of NCI by $1,000$ implies a reduction of $0.12$ to $0.15$ in $R_{I}$ and $0.06$ to $0.09$ in $R_{II}$, which is generally in line with the expectation that the measures capture a distance to default.

\begin{table}[ht]
\centering
\caption{Determinants of $ R_{I}$ and $ R_{II}$ with NCI index, controlling for financial markets and macroeconomic factors} \label{table5x}
\includegraphics[scale=.6]{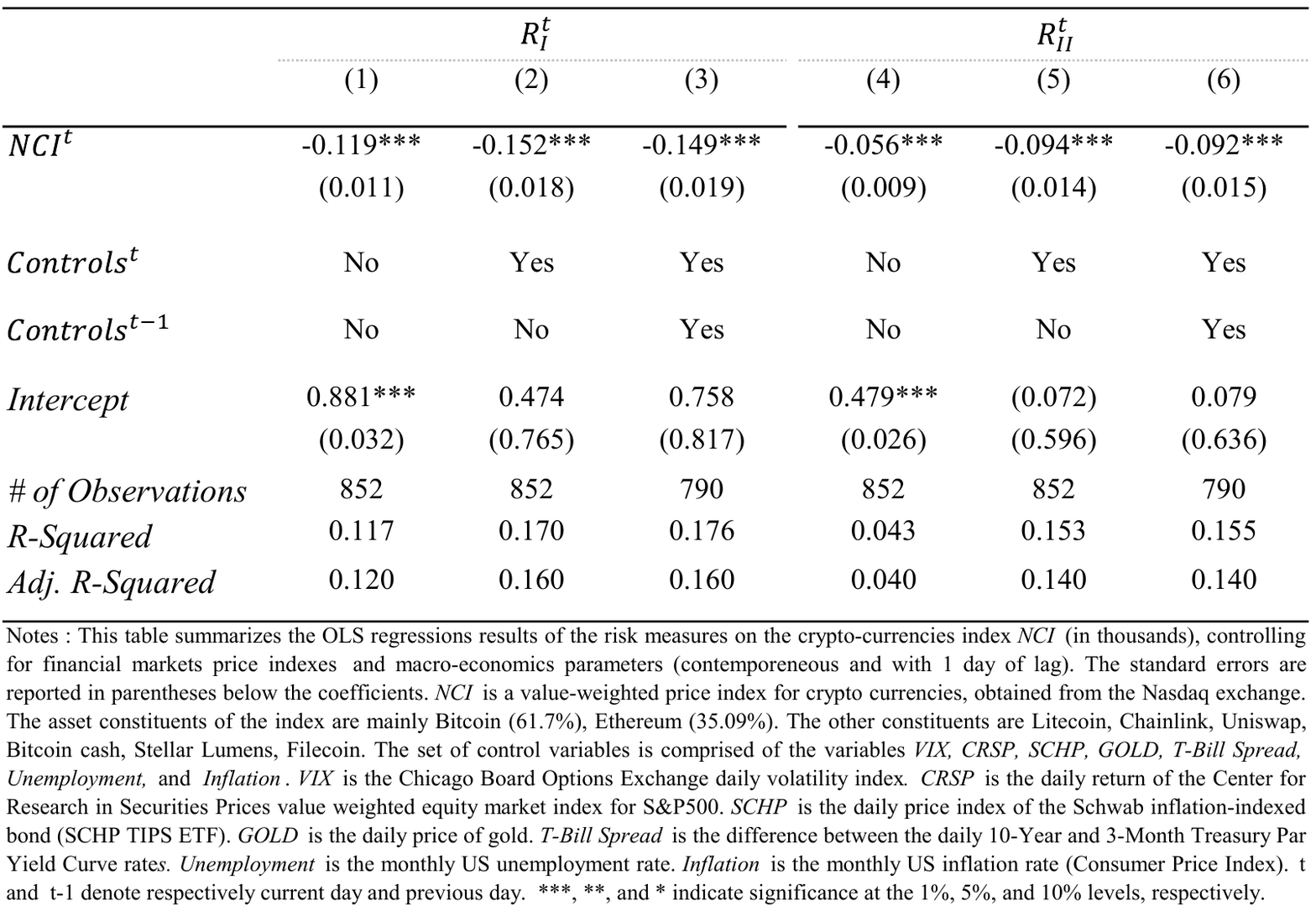}
\end{table}

To provide further evidence, we examine the statistical relationship between the risk measures and liquidations. Note that liquidations may not necessarily occur during normal market conditions, because a borrower is often better-off  closing the position before it becomes eligible for liquidation, over paying the bonus to a liquidator: in this respect, liquidations under DeFi are more costly than a traditional margin call because liquidators must be incentivized. Nevertheless, liquidations are a proxy for elevated risks in the system. \

\begin{figure}[ht]
\begin{center}
\includegraphics[width=1\linewidth]{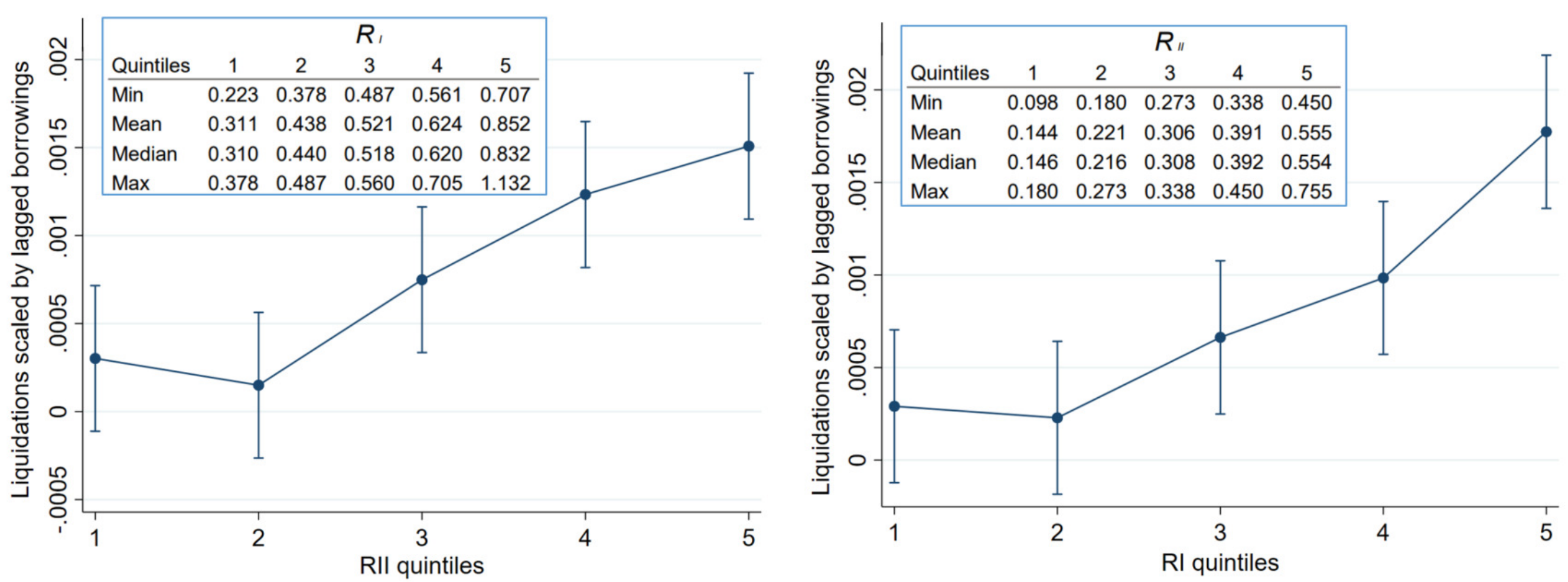}
\caption{Liquidations scaled by lagged borrowings - Quintiles of $R_I$ (left), $R_{II}$ (right)}\label{fig6x}
\end{center}
\label{fig11}
\end{figure}

Given that risk measures are invariant to the size of the market, we similarly define liquidations as the ratio of dollar liquidations to lagged dollar borrowings. Risk measures and liquidations are computed separately for AAve and Compound from blockchain queries.\footnote{The query is available at https://dune.com/queries/463010.} We winsorize the liquidations at the 1st and 99th percentile levels to reduce the effect of outliers. Figure \ref{fig11} reveals that, overall, liquidations are increasing in the risk measure, consistent with our measures capturing risk at an aggregated level. 

\FloatBarrier

\section*{Conclusion}

\paragraph{}
Shortly following critical Fintech innovations in automated lending and payment systems \cite{das19,agarwalzhang20,allengujagtiani22}, decentralized finance offers new opportunities for direct matching of borrowers and lenders \cite{harramsan21}, but also introduces new risks that are no longer centrally monitored. In this study, we offer a practical measure of DeFi lending risk based on a decomposition of aggregate lending and borrowing into a synthetic borrower and a synthetic lender. The measure can be obtained without knowledge of individual net positions, and is interpretable in terms of maximal allowable changes in cryptocurrency prices held by borrowers. We apply the measure to detect significant increases in risk identifying in real time potential fragilities in the system. The measure offers a first step to track risks at an aggregate level and is based on publicly available custom queries on the blockchain that, we hope, can lead to further research to construct real-time assessments of system risk. Our analysis leaves open other important aspects of this sector, such as, given a long enough time-series, obtaining a more systematic empirical understanding of financial risks as previously examined for other cryptoassets \cite{liutsy21,liutsywu22,borliutsy22}.

In the event of recurrent high risk level, the evolution of the industry may be reflecting a movement from guaranteed deposits to fiat banking, similar to a traditional bank and sensitive to bank runs \'{a} la \citeasnoun{diadyb83}. On its own, this evolution, if matched with higher interest rates, may reflect economic demands for higher yields but comes with new risks that are not yet fully understood nor monitored. Indeed, many of the safeguards currently in place to protect against under-collateralized lending, such as issuing the platform's governance coin (e.g., the Compound coin), are similar to those in place in the recent Terra USD coin collapse, which called for issuing the Luna coin to meet any excess withdrawal. The new banking system proposed in DeFi has become more sensitive to loss of trust in the protocol and its associated governance coin. While we have not explored these questions here, these challenges call for a complete economic theory of DeFi lending where the incentives of lenders, borrowers, liquidators and sponsors are better understood. 

\end{doublespace}
\ifx\undefined\bysame
\newcommand{\bysame}{\leavevmode\hbox to\leftmargin{\hrulefill\,\,}}
\fi

\end{document}